\documentclass{article}

\usepackage{amssymb}
\usepackage{amsmath}
\usepackage{latexsym}
\usepackage[pdftex]{graphicx}
\usepackage{caption}
\usepackage{subcaption}
\usepackage{url}
\usepackage{cancel}
\usepackage{soul}
\usepackage{paralist}
\usepackage{fancybox}
\usepackage[ruled,vlined]{algorithm2e}	
\usepackage{nccrules}
\usepackage{dashrule}
\usepackage{bm}
\usepackage{stackrel}
\usepackage{color}
\usepackage[dvipsnames]{xcolor}
\usepackage[scr=boondoxo, scrscaled=.98]{mathalfa}
\usepackage{authblk}
\usepackage{hyperref}

\newcommand{\A}{\mathcal{A}} 

\newcommand{\E}{\mathcal{E}}

\newcommand{\J}{\mathcal{J}}

\newcommand{\N}{\mathcal{N}}
 \renewcommand{\P}{\mathcal{P}}

\newcommand{\T}{\mathcal{T}}

\newtheorem{theorem}{Theorem}
\newtheorem{proposition}{Proposition}
\newtheorem{definition}{Definition}
\newtheorem{example}[theorem]{Example}
\newtheorem{problem}[theorem]{Problem}

\newtheorem{exampleAux}{Example}%
\renewenvironment{example}{\begin{exampleAux}\upshape}%
{\hfill\qedbox\end{exampleAux}}

\newenvironment{proof}[1][Proof]{\begin{trivlist}
\item[\hskip \labelsep {\itshape #1.}]}{\qed\end{trivlist}}

\newenvironment{exampleCont}[1]{\trivlist
  \item[\hskip \labelsep{\textit{Example~#1 (cont.).}}]}{\markfull\endtrivlist}

\newlength{\qedlengte}
\newcommand{\qedbox}{\rule{\qedlengte}{\qedlengte}}

\def\qed{\hfill{\qedboxempty}     
  \ifdim\lastskip<\medskipamount \removelastskip\penalty55\medskip\fi}

\def\qedboxempty{\vbox{\hrule\hbox{\vrule\kern3pt
                 \vbox{\kern3pt\kern3pt}\kern3pt\vrule}\hrule}}

\def\qedfull{\hfill{\qedboxfull}  
  \ifdim\lastskip<\medskipamount \removelastskip\penalty55\medskip\fi}

\def\qedboxfull{\vrule height 4pt width 4pt depth 0pt}

\newcommand{\markfull}{\qedfull}

\newcommand{\uD}{\mathcal{D}}

\newcommand{\C}{{\cal C}}

\newcommand{\set}[1]{\{#1\}}

\newcommand{\D}{\textbf{D}}

\newcommand{\I}{\mathcal{I}}

\renewcommand{\r}{\textit{r}}

\newcommand{\ri}{\r^{\I}}
\renewcommand{\S}{\mathcal{S}}

\newcommand{\dom}[1]{\mathit{dom}(#1)}
\newcommand{\tup}[1]{\langle #1 \rangle}

\newcommand{\inp}{{ i}}
\newcommand{\outp}{{ o}}
\newcommand{\inputmode}[1]{#1^{\inp}}
\newcommand{\pattern}{\mathrm{\Pi}}
\newcommand{\patternSet}{\boldsymbol{\mathrm{\Pi}}}

\newcommand{\accessPathSet}{\boldsymbol{P}}

\newcommand{\reach}{\mathit{reach}}

\newcommand{\arc}{\mbox{$^{\curvearrowright}$}}
\newcommand{\dgraph}[2]{G_{#1}^{#2}}
\newcommand{\GqS}{\dgraph{q}{\SP}}
\newcommand{\SP}{\S^{\patternSet}}

\newcommand{\access}[3]{\overset{#1}{\longrightarrow_{#2^{\I}}}#3}
\newcommand{\ProgramSet}{\mathbf{P}^{\SP}_{q}}
\newcommand{\Program}{P}
\newcommand{\Accesses}[2]{Acc(#1,#2)}
\newcommand{\SJ}{\J}
\newcommand{\binding}{\mathscr{b}}

\newcommand{\sel}{\mbox{\large$\sigma$}}      
\newcommand{\proj}{\mbox{\large$\pi$}}        
\newcommand{\join}{\mbox{\large$\bowtie$}}

\def\codeif{\mbox{\upshape\textbf{if}}}
\def\codethen{\mbox{\upshape\textbf{then}}}
\def\codefor{\mbox{\upshape\textbf{for}}}
\def\codelet{\mbox{\upshape\textbf{let}}}
\def\codeforeach{\mbox{\upshape\textbf{for each}}}

\def\codewhile{\mbox{\upshape\textbf{while}}}
\def\codetrue{\mbox{\upshape\textbf{true}}}
\def\codefalse{\mbox{\upshape\textbf{false}}}
\def\codereturn{\mbox{\upshape\textbf{return}}}

\def\codeelse{\mbox{\upshape\textbf{else}}}
\def\codenil{\mbox{\upshape\textbf{nil}}}
\def\codeexit{\mbox{\upshape\textbf{exit}}}

\newcommand{\ind}{\dashrule[-0.6ex]{0.4}{1 1 1 1 1 1 1 1 1 1 1 1}\quad}

\newcommand{\compatible}{\texttt{compatible}}
\newcommand{\answerable}{\texttt{answerable}}
\newcommand{\connected}{\texttt{connected}}
\newcommand{\peel}{\texttt{peel}}
\newcommand{\extract}{\texttt{\upshape extract}}

\newcommand{\followPath}{\texttt{followPathDepthFirst}}

\newif\ifistechrep
\istechreptrue

\begin{document}

\title{Keyword Search in the Deep Web}

\author{Andrea Cal\`i$^{1}$ \and Davide Martinenghi$^2$ \and Riccardo
  Torlone$^3$}

\urldef{\autone}\url{emilio.delorenzis@mail.polimi.it}
\urldef{\auttwo}\url{davide.martinenghi@polimi.it}

\affil{$^1$Birkbeck, University of London, UK,\quad \texttt{andrea{@}dcs.bbk.ac.uk}\\
$^2$Politecnico di Milano, Italy,\quad \texttt{davide.martinenghi{@}polimi.it}\\
$^3$Universit\`a Roma Tre, Italy,\quad \texttt{torlone{@}dia.uniroma3.it}\\
}
 
\date{}

\pagestyle{plain} 

\sloppy

\maketitle

\begin{abstract}
  The Deep Web is constituted by data that are
  accessible through Web pages, but not readily 
  indexable by search engines
  as they are returned in dynamic pages.  In this paper we propose a conceptual framework for answering keyword queries on Deep Web sources 
  represented as relational tables with so-called access limitations. We formalize the notion of optimal answer, characterize queries for which an answer can be found, 
  and present a method for query processing based on the construction of a query plan that minimizes the accesses to the data sources.
\end{abstract}

\section{Introduction}

It is well known that the portion of the Web indexed by search engines constitutes only a very small fraction of the data available online. The vast majority of the data, commonly referred to as \emph{Deep Web}, is ``hidden'' in local databases whose content can only be accessed by manually filling up Web forms.
This happens for instance when we need to find a flight from Italy to Japan on the Web site of an airline company. This immediately poses an interesting challenge, i.e., how to automatically retrieve relevant information from the Deep Web -- a problem that has been deeply investigated in recent years (see, e.g.,~\cite{BiMi07,CaMa10,MAAH09,DBLP:conf/vlds/BienvenuDMSS12} for discussion).

Usually, a data source in the Deep Web is conceptually modeled by a relational table in which some columns, called \emph{input} attributes, represent fields of a form that need to be filled in so as to retrieve data from the source, while all the others, called \emph{output} attributes, represent values that are returned to the user. Consider for instance the following relations in which the $i$ superscript denotes the input attributes.
\begin{center}
$r_1=$
    \begin{tabular}{|cc|l} \cline{1-2}
	   $\inputmode{\textit{Dept}}$ & \textit{Emp} \\
	  \cline{1-2}
	IT & John & $t_{11}$\\
	AI & Mike & $t_{12}$\\
	  \cline{1-2}
	\end{tabular}
    \quad
$r_2=$
    \begin{tabular}{|cc|l} \cline{1-2}
	  $\inputmode{\textit{Emp}}$ & \textit{Proj} \\
	  \cline{1-2}
	John & P1 & $t_{21}$\\
	Ann & P2 & $t_{22}$\\
	Mike & P2 & $t_{23}$\\
	  \cline{1-2}
	\end{tabular}
    \quad
$r_3=$
    \begin{tabular}{|ccc|l} \cline{1-3}
	 $\inputmode{\textit{Proj}}$ & \textit{Emp} & \textit{Role} \\	  
	 \cline{1-3}
	P1 & John & DBA & $t_{31}$\\
	P1 & Ann & Analyst & $t_{32}$\\
	  \cline{1-3}
	\end{tabular}
\end{center}
Relation $r_1$ represents a form that, given a department, returns all the employees working in it; relation $r_2$ a form that, given an employee, returns all the projects he/she works on; and relation $r_3$ a form that, given a project, returns the employees working in it along with their role. These access modalities are commonly referred to as \emph{access limitations}, in that data can only be queried according to given patterns.

Different approaches have been proposed in the literature for querying databases with access limitations: conjunctive queries~\cite{CaMa:ICDE2008}, natural language~\cite{Lehm12}, and SQL-like statements~\cite{JaJa15}. In this paper, we address the novel problem of accessing the Deep Web by just providing a set of keywords, in the same way in which we usually search for information 
 on the Web with a search engine. 

Consider for instance the case in which the user only provides the keywords
``DBA''  and ``IT'' for querying the portion of the Deep Web represented by the relations above. Intuitively, he/she is searching for employees with the DBA role in the IT department. Given the access limitations, this query can be concretely answered by first accessing relation $r_1$ using the keyword  \emph{IT}, which allows us to extract the tuple $t_{11}$. Then, using the value \emph{John} in $t_{11}$, we can extract the tuple $t_{21}$ from relation $r_2$. Finally, using the value \emph{P1} in $t_{21}$, we can extract the tuples $t_{31}$ and $t_{32}$ from relation $r_3$. Now, since $t_{31}$ contains \emph{DBA}, it turns out that the set of tuples $\{t_{11},  t_{21}, t_{31}\}$ is a possible answer to the input query in that the set is connected (every two tuples in it share a constant) and contains the given keywords. 
However, 
 the tuple $t_{21}$ is somehow redundant and can be safely eliminated from the solution, since the set $\{t_{11}, t_{31}\}$ is also connected and contains the keywords. This example shows that, in this context, the keyword query answering problem
 can be involved and tricky, even in simple situations. 

In the rest of this paper, we formally investigate this problem in depth. We first propose, in Section~\ref{sec:problem}, a precise semantics of (optimal) answer to a keyword query in the Deep Web. We then tackle, in Section~\ref{sec:tsearch},  the problem of finding an answer to a keyword query by assuming that the domains of the keywords are known in advance. This allows us to perform static analysis to immediately discard irrelevant cases from our consideration. In this framework, in Section~\ref{sec:extraction} we introduce the notion of minimal query plan aimed at efficiently retrieving an answer to a keyword query and, in Section~\ref{sec:plan-generation}, we present a technique for its automatic generation that can be generalized to cases in which the domains of the keyword are unknown.
Section~\ref{sec:related} discusses related work.
Section~\ref{sec:conclusions} ends the paper with some conclusions and future works.
All proofs of our claims are available in 
\ifistechrep
the appendix.
\else
a technical report~\cite{MT2016:techrep}.
\fi

This technical report is the extended version of~\cite{DBLP:conf/er/CaliMT16}.

\section{Preliminaries and problem definition}
\label{sec:problem}

We model data sources as relations of a relational
database and we assume that, albeit autonomous, they have ``compatible''
attributes. For this, we fix a set of
\emph{abstract domains} $\D=\{D_1,\ldots,D_m\}$, which, rather
than denoting concrete value types (such as string or integer), represent
data types at a higher level of abstraction (for instance, \emph{car} or
\emph{country}). Therefore, in an abstract domain an object is uniquely represented by a value.
The set of all values is denoted by $\uD = \bigcup_{i=1}^{n} D_i$.
For simplicity, we assume that all abstract domains are disjoint.
We then say that a \emph{(relation) schema} $\r$, customarily indicated as $\r(A_1,\ldots,A_k)$, is a set of attributes $\set{A_1, \ldots, A_k}$, each
associated with an abstract domain $\dom{A_i}\in \D$, $1 \leq i \leq k$.  A
\emph{database schema} $\S$ is a set of
\emph{schemas} $\set{\r_1, \ldots, \r_n}$.

As usual, given a schema $\r$, a \emph{tuple} $t$ over $\r$ is a function that
associates a value $c\in \dom{A}$ with each attribute $A\in \r$, and a
\emph{relation} \emph{instance} $\ri$ of $\r$ is a set of tuples over $\r$.
For simplicity, we also write $\dom{c}$ to indicate the domain of $c$.
A (database) instance $\I$ of a database schema $\S = \set{\r_1, \ldots, \r_n}$
is a set of relation instances
$\set{\ri_1, \ldots, \ri_n}$, where $\ri_i$ denotes the relation instance of $\r_i$ in $\I$.

For the sake of simplicity, in the following we assign the same name to attributes of different schemas that are defined over the same abstract domain.

\begin{definition}[Access pattern]
An \emph{access pattern} $\pattern$ for a schema
$\r(A_1,\ldots,A_k)$ is a mapping $\pattern: \{A_1,\ldots,A_k\}\rightarrow M$, where $M=\{\inp,\outp\}$ is called \emph{access mode}, and
$\inp$ and $\outp$ denote \emph{input} and \emph{output}, respectively; $A_i$ is correspondingly called an \emph{input} (resp., \emph{output}) \emph{attribute} for $\r$ wrt $\pattern$.
\end{definition}
Henceforth, we denote input attributes with an `$\inp$' superscript, e.g., $\inputmode{A}$.
Moreover, we assume that each relation has exactly one access pattern.

\begin{definition}[Binding]
Let $A'_1,\ldots, A'_\ell$ be all the input attributes for $\r$ wrt $\pattern$; any tuple $\binding=\langle c_1, \ldots, c_\ell\rangle$ such that $c_i\in\dom{A'_i}$ for $1\leq i\leq \ell$ is called a \emph{binding} for $\r$ wrt $\pattern$.
\end{definition}
\begin{definition}[Access]
An \emph{access} is a pair $\langle\pattern,\binding\rangle$, where $\pattern$ is an access pattern for a schema $\r$ and $\binding$ is a binding for $\r$ wrt $\pattern$. The \emph{output} of such an access on an instance $\I$ is the set $\T$ of all tuples in the relation $\ri\in\I$ over $\r$ that match the binding, i.e., such that
$\T=\sel_{A_1=c_1,\ldots, A_\ell=c_\ell}(r).$
\end{definition}
Intuitively, we can only access a relation if we can provide a binding for it, i.e., a value for every input attribute.

\begin{definition}[Access path]\label{def:access-path}
Given an instance $\I$ for a database schema $\S$, a set of access patterns $\patternSet$ for the relations in $\S$, and a set of values $\C\subseteq\uD$,
an \emph{access path} on $\I$ (for $\S$, $\patternSet$ and $\C$) is a sequence
$\access{\binding_1}{r_1}{\T_1}\access{\binding_2}{r_2}{}\cdots\access{\binding_n}{r_n}{\T_n}$,
where, for $1\leq i\leq n$,
\begin{inparaenum}[\itshape (i)]
	\item $\binding_i$ is a binding for a relation $\r_i\in\S$ wrt a pattern $\pattern_i\in\patternSet$ for $\r_i$,
	\item $\T_i$ is the output of access $\langle \pattern_i,\binding_i\rangle$ on $\I$, and
	\item each value in $\binding_i$ either occurs in $\T_j$ with $j<i$ or is a value in $\C$.
\end{inparaenum}
\end{definition}

\begin{figure}
  \centering
  \begin{subfigure}[b]{0.43\textwidth}
      \centering
	  \includegraphics[width=.87\textwidth]{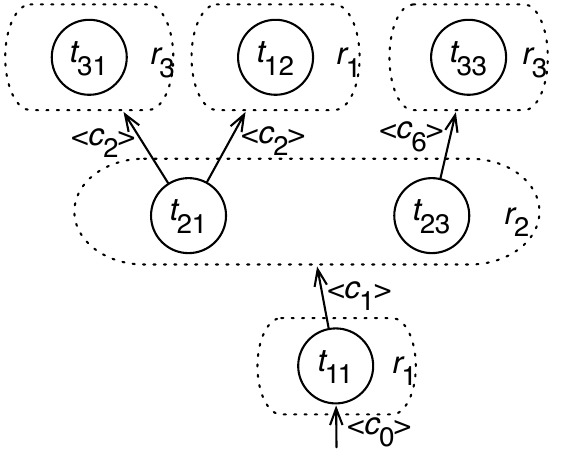}
\caption{Reachable portion $\I'$, given $\{c_0\}$.}
	\label{fig:accessPaths}
  \end{subfigure}
  \begin{subfigure}[b]{0.25\textwidth}
      \centering
	  \includegraphics[width=0.947\textwidth]{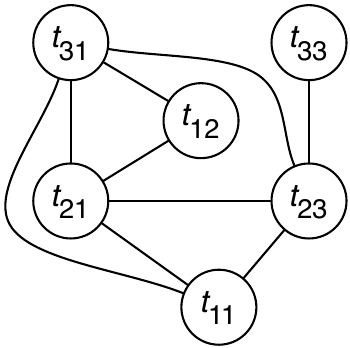}
\caption{Join graph of $\I'$.}
	\label{fig:joinGraph}
  \end{subfigure}
  \begin{subfigure}[b]{0.3\textwidth}
      \centering
	  \includegraphics[width=1.15\textwidth]{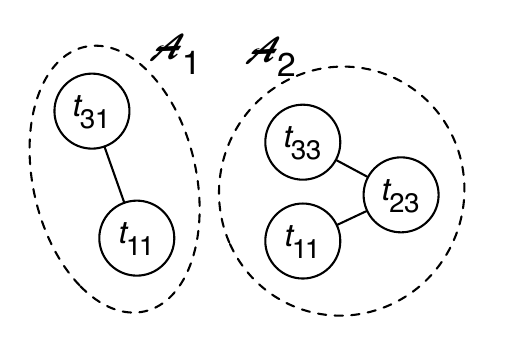}
\caption{Two answers to KQ $q$.}
	  	\label{fig:answers}
  \end{subfigure}
  \caption{Illustration of Examples~\ref{ex:reachable}, \ref{ex:join-graph}, and~\ref{ex:answer}.}
  \label{fig:initial-example}
\end{figure}

\begin{definition}[Reachable portion]
A tuple $t$ in $\I$
is said to be \emph{reachable} given $\C$ if there exists an access path $P$
(for $\S$, $\patternSet$ and $\C$)
such that $t$ is in the output of some access in $P$;
the \emph{reachable portion} $\reach(\I,\patternSet,\C)$ of $\I$ is the set of all reachable tuples in $\I$ given $\C$.
\end{definition}
In the following, we will write $\SP$ to refer to schema $\S$ under access patterns $\patternSet$.
\begin{example}\label{ex:reachable}
Consider the following instance $\I$ of a schema $\SP=\{\r_1(\inputmode{A_1},A_2), \r_2(\inputmode{A_2},A_1), \r_3(\inputmode{A_1},A_2,A_3)\}$.
	\begin{center}
		$r_1=$
    \begin{tabular}{|cc|l} \cline{1-2}
	   $\inputmode{A_1}$ & $A_2$ \\
	  \cline{1-2}
	$c_0$ & $c_1$ & $t_{11}$\\
	$c_2$ & $c_3$ & $t_{12}$\\
	  \cline{1-2}
	\end{tabular}
    \quad
	$r_2=$
    \begin{tabular}{|cc|l} \cline{1-2}
	   $\inputmode{A_2}$ & $A_1$ \\
	  \cline{1-2}
	$c_1$ & $c_2$ & $t_{21}$\\
	$c_4$ & $c_2$ & $t_{22}$\\
	$c_1$ & $c_6$ & $t_{23}$\\
	  \cline{1-2}
	\end{tabular}
    \quad
	$r_3=$
    \begin{tabular}{|ccc|l} \cline{1-3}
	   $\inputmode{A_1}$ & $A_2$ & $A_3$ \\
	  \cline{1-3}
	$c_2$ & $c_1$ & $c_8$ & $t_{31}$\\
	$c_5$ & $c_4$ & $c_8$ & $t_{32}$\\
	$c_6$ & $c_8$ & $c_9$ & $t_{33}$\\
	  \cline{1-3}
	\end{tabular}
	\end{center}
Then, for instance, $\{t_{11}\}$ is the output of the access
with binding $\langle c_0\rangle$ wrt $r_1(\inputmode{A_1},A_2)$, and 
$\access{\tup{c_0}}{r_1}{\{t_{11}\}}\access{\tup{c_1}}{r_2}{\{t_{21},t_{23}\}}$,
is an access path for  $\S$, $\patternSet$ and $\C=\{c_0\}$, since, given $\C$, we can extract $t_{11}$ from $r_1$ and, given $\{c_1\}$ from $t_{11}$, we can extract $t_{21}$ and $t_{23}$ from $r_2$. The reachable portion of $\I$, given $\C$, is $\reach(\I,\patternSet,\C)=\{t_{11}, t_{12}, t_{21}, t_{23}, t_{31}, t_{33}\}$, while $\{t_{22}, t_{32}\}\cap \reach(\I,\patternSet,\C)=\emptyset$. Figure~\ref{fig:accessPaths} shows the reachable portion $\I'$ of $\I$ given $\C$ along with the access paths used to extract it, with dotted lines enclosing outputs of accesses.
\end{example}

\noindent
The definition of answer to a keyword query in our setting requires the preliminary notion of join graph.
\begin{definition}[Join graph]
Given a set $\T$ of tuples, the \emph{join graph} of $\T$ is a node-labeled
undirected graph $\tup{N,E}$ constructed as follows:
\begin{inparaenum}[\itshape (i)]
\item the nodes $N$ are labeled with tuples of $\T$, with a one-to-one correspondence between tuples of $\T$ and nodes of $N$; and
\item there is an arc between two nodes $n_1$ and $n_2$ whenever the tuples labeling
  $n_1$ and $n_2$ have at least one value in common.
  \end{inparaenum}
\end{definition}

\begin{example}\label{ex:join-graph}
Consider instance $\I$ of Example~\ref{ex:reachable} and the reachable portion $\I'$ of $\I$ given $\{c_0\}$, shown in Figure~\ref{fig:accessPaths}. The join graph of $\I'$ is shown in Figure~\ref{fig:joinGraph}.
\end{example}
A \emph{keyword query} (KQ) is a non-empty set of values in $\uD$ called \emph{keywords}.
\begin{definition}[Answer to a KQ]\label{def:answer}
  An answer to a KQ $q$ against a database instance $\I$ over a
  schema $\SP$ is a set of tuples $\A$ in
  $\reach(\I,\patternSet,q)$ such that:
  \begin{inparaenum}[\itshape (i)]
  \item each keyword $k \in q$ occurs in at least one tuple $t$ in $\A$;
  \item the join graph of $\A$ is connected;
  \item no proper subset $\A' \subset \A$ satisfies both Conditions~(i) and~(ii) above.
  \end{inparaenum}
\end{definition}
It is straightforward to
see that there could be several answers to a KQ;
below we give a widely accepted criterion for ranking such answers~\cite{XuQC10}.

\begin{definition} \label{def:optimal}
  Let $\A_1, \A_2$ be two answers to a KQ $q$ on an instance $\I$.
  We say that $\A_1$ \emph{is better than}
  $\A_2$
  if
  $|\A_1|\leq|\A_2|$.
  The \emph{optimal} answers are those of minimum size.
\end{definition}

\begin{example}\label{ex:answer}
Consider a KQ $q=\{c_1,c_8\}$ over the instance $\I$ of Example~\ref{ex:reachable}.
Figure~\ref{fig:accessPaths} shows two possible answers: $\A_1 = \{t_{11}, t_{31}\}$ and $\A_2= \{t_{11}, t_{23}, t_{33}\}$. $\A_1$ is better than $\A_2$ and is the optimal answer to $q$.
\end{example}
In the following sections we will tackle the following problem.
\begin{problem}\label{prob:problem}
	To efficiently find an (optimal) answer to a KQ.
\end{problem}

\section{Detecting non-answerable queries}
\label{sec:tsearch}

In this section, we tackle Problem~\ref{prob:problem} by assuming that the domains of the keywords are known in advance.
This allows us to perform static analysis to immediately discard irrelevant cases from our consideration.
We will later show how to extend our techniques to cover the cases in which the domains of the keywords are not known.
For convenience of notation, we sometimes write $c\colon D$ to denote value $c$ and indicate that $\dom{c}=D$.
In addition, in our examples, the name of an attribute will also indicate its abstract domain.
\subsection{Compatible queries}
In order to focus on meaningful queries, in this and in the following subsection, we semantically characterize queries for which an answer might be found.
\begin{definition}[Compatibility]\label{def:compatible}
A KQ $q$ is said to be \emph{compatible} with a schema $\S$
if there exist a set of access patterns $\patternSet$ and
 an instance $\I$ over $\SP$
such that 
there is an answer to $q$ against $\I$.
\end{definition}
\begin{example}\label{ex:compatibility}
The KQ $q_1=\{a\colon A,c\colon C\}$ is not compatible with schema $\S_1=\{r_1(A,B),r_2(C,D)\}$, since no set of tuples from $\S$ containing all the keywords in $q_1$ can ever be connected, independently of the access patterns for $\S_1$. Conversely, $q_1$ is compatible with $\S_2=\{r_1(A,B),r_3(B,C)\}$, as witnessed by a possible answer $\{r_1(a,b),r_3(b,c)\}$ and patterns $\patternSet$ such that $\SP_2=\{r_1(\inputmode{A},B),r_3(B,C)\}$.

Similarly, KQ $q_2=\{a\colon A,a'\colon A\}$ is compatible with a schema $\S_3=\{r_1(A,B)\}$, as witnessed by a possible answer $\{r_1(a,b),r_1(a',b)\}$ and patterns $\patternSet$ such that $\SP_3=\{r_1(\inputmode{A},B)\}$. However, $q_2$ is not compatible with a schema $\S_4=\{r_4(A)\}$, since a unary relation, alone, can never connect two keywords.
\end{example}
In order to check compatibility of a KQ with a schema, we refer to the notion of schema join graph and then proceed as described in Algorithm~\ref{alg:compatibility}.
\begin{definition}[Schema join graph]
Given a schema $\S$, the \emph{schema join graph} of $\S$ is a node-labeled
undirected graph $\tup{\N,E}$ constructed as follows:
\begin{inparaenum}[\itshape (i)]
\item the nodes $\N$ are labeled with relations of $\S$, with a one-to-one correspondence between relations of $\S$ and nodes of $\N$; and
\item there is an arc between two nodes $N_1$ and $N_2$ in $\N$ (including self-loops) whenever the relations labeling $N_1$ and $N_2$ have at least one attribute with the same domain.
\end{inparaenum}
\end{definition}
\begin{algorithm}[ht]
\scalebox{.82}{
		\begin{minipage}{1.33\textwidth}
				\begin{compactenum}
				    \item[Input:] \emph{Schema $\S$, KQ $q=\{k_1,\ldots,k_{|q|}\}$}
				    \item[Output:] \emph{$\codetrue$ if $q$ is compatible with $\S$, $\codefalse$ otherwise}
					\item\label{algline:domainnotfound} $\codeif$ $\exists k_i\in q$ s.t. no attribute in $\S$ has domain $\dom{k_i}$
					$\codethen$ $\codereturn$ $\codefalse$
					\item\label{algline:nonunary} $\SJ := $ schema join graph of $\S$ excluding the relations of arity $1$
					\item $\codefor$ $i$ in $1...|q|-1$
					\item\label{algline:ifnopath} \ind $\codeif$ $\exists$ no path in $\SJ$ connecting two attributes with domains $\dom{k_i}$ and  $\dom{k_{i+1}}$
					\item \ind\ind $\codereturn$ $\codefalse$
					\item $\codereturn$ $\codetrue$
				\end{compactenum}
        \end{minipage}
        }
	\caption{Checking compatibility ($\compatible(q,\S)$)}
	\label{alg:compatibility}
\end{algorithm}
The main idea behind Algorithm~\ref{alg:compatibility} is that the domains of any two keywords in the KQ must be properly connected in the schema join graph. In order for an answer to ever be possible, we must find an instance that exhibits a witness (i.e., a set of tuples) satisfying all the conditions of Definition~\ref{def:answer}.
\begin{proposition}\label{pro:compatibility-correct}
Algorithm~\ref{alg:compatibility} correctly checks compatibility in PTIME.
\end{proposition}
\subsection{Answerable queries}
\begin{figure}
  \centering
  \begin{subfigure}[b]{0.45\textwidth}
      \centering
	  \includegraphics[width=.9\textwidth]{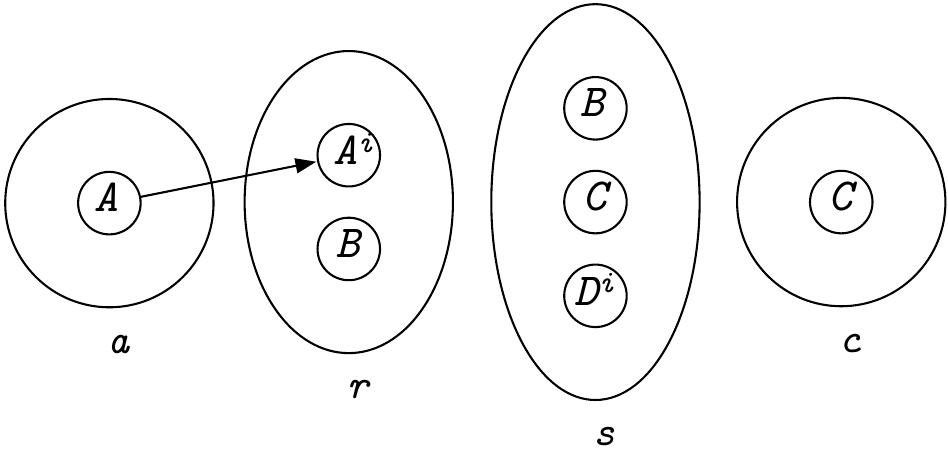}
  \caption{D-graph $\dgraph{q}{\SP_1}$ from Example~\ref{ex:answerability}; $s$ is not visible.}
      \label{fig:answerability1}
  \end{subfigure}
\hfill
  \begin{subfigure}[b]{0.45\textwidth}
      \centering
	  \includegraphics[width=.9\textwidth]{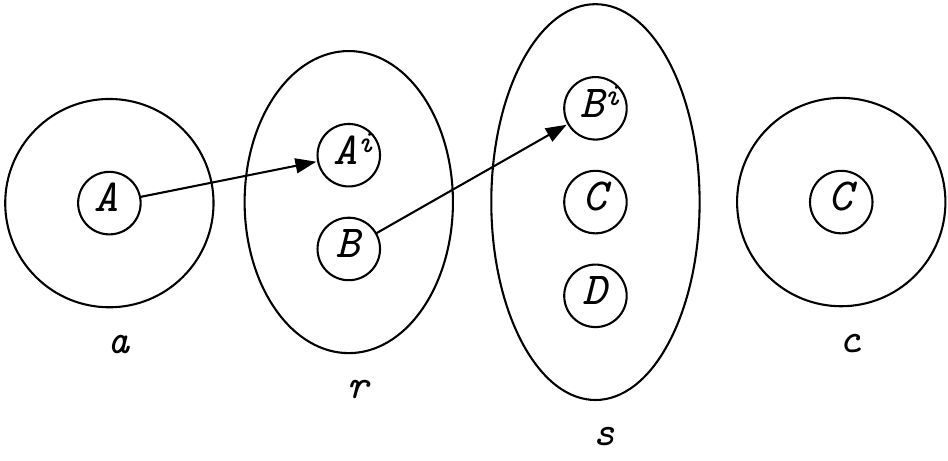}
  \caption{D-graph $\dgraph{q}{\SP_2}$ from Example~\ref{ex:answerability}; all relations are visible.}
      \label{fig:answerability2}
  \end{subfigure}
  \begin{subfigure}[b]{0.45\textwidth}
      \centering
	  \includegraphics[width=\textwidth]{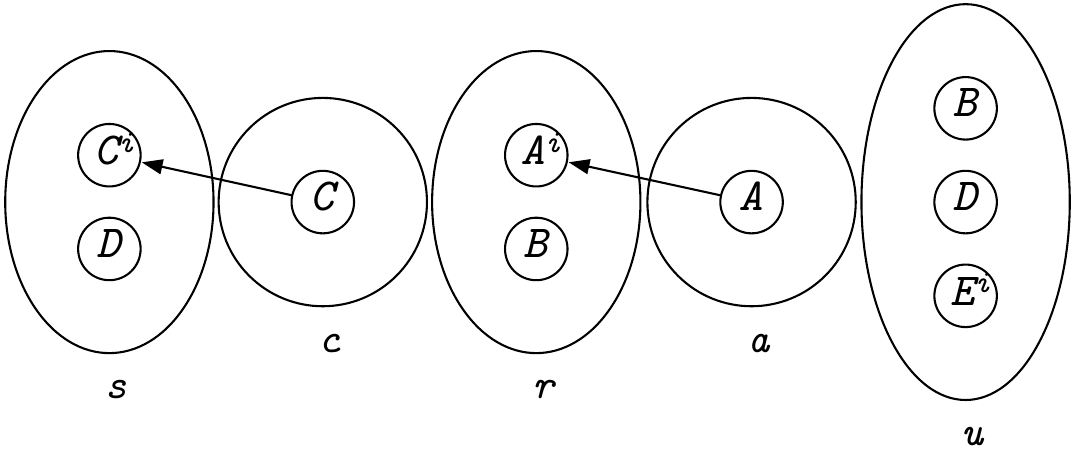}
  \caption{D-graph $\dgraph{q}{\SP}$ from Example~\ref{ex:nonanswerable}; $u$ is not visible.}
      \label{fig:nonanswerable}
  \end{subfigure}
\hfill  
  \begin{subfigure}[b]{0.45\textwidth}
      \centering
	  \includegraphics[width=.9\textwidth]{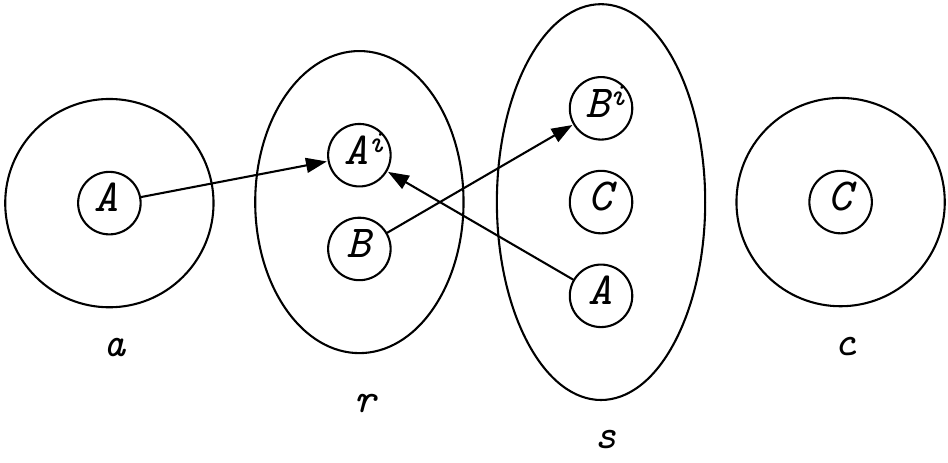}
  \caption{D-graph $\dgraph{q}{\SP}$ from Example~\ref{ex:cycle-extraction} with cyclic chain of relations $r-s$.}
      \label{fig:cyclic-dgraph}
  \end{subfigure}
  \caption{D-graphs from the Examples.}
  \label{fig:dgraphs}
\end{figure}

A stricter requirement than compatibility is given by the notion of answerability.
\begin{definition}[Answerability]\label{def:answerable}
A KQ $q$ is \emph{answerable} against a schema $\SP$
if there is an instance $\I$ over $\SP$
such that there is an answer to $q$ against $\I$.
\end{definition}
\begin{example}\label{ex:answerability}
Consider KQ $q=\{a\colon A,c\colon C\}$ and schema $\SP_1=\{r(\inputmode{A},B),$ $s(B,C,\inputmode{D})\}$.
Although $q$ is compatible with $\S_1$, it is not answerable against $\SP_1$, since no tuple from $\S_1$ can be extracted under $\patternSet$ (no values for domain $D$ are available).
Conversely, $q$ is answerable against $\SP_2=\{r(\inputmode{A},B),s(\inputmode{B},C,D)\}$, since an answer like $\{r(a,b),s(b,c,d)\}$ could be extracted by first accessing $r$ with binding $\tup{a}$, thus extracting value $b$, and then $s$ with binding $\tup{b}$.
\end{example}
In order to check answerability, we need to check that all the required relations can be accessed according to the access patterns.
To this end,
we first refer to a schema enriched with unary relations representing the keywords in the KQ.
\footnote{If other values are known besides the keywords, this knowledge may be represented by means of appropriate unary relations with output mode in the schema.}
\begin{definition}[Expanded schema]\label{def:expandedSchema}
Let $q$ be a KQ over a schema $\SP$.
The \emph{expanded schema} $\SP_q$ of $\SP$ wrt. $q$ is defined as $\SP_q=\SP \cup \{r_c(C)|c\in q\}$, where $r_c$ is a new unary relation, not occurring in $\SP$, whose only attribute $C$ is an output attribute with abstract domain $\dom{C}=\dom{c}$.
\end{definition}
Then, we use the notion of dependency graph (d-graph) to denote output-input dependencies between relation arguments, indicating that a relation under access patterns needs values from other relations.
\begin{definition}[d-graph]\label{def:d-graph}
Let $q$ be a KQ over a schema $\SP$.
The \emph{d-graph} $\dgraph{q}{\SP}$
is a directed graph $\tup{\N,\E}$ defined as follows.
For each attribute $A$ of each relation in the expanded schema $\SP_q$, there is a node in $\N$ labeled with $A$'s access mode and abstract domain.
There is an arc $u\arc v$ in $\E$ whenever: (\textit{i}) $u$ and $v$ have the same
abstract domain; (\textit{ii}) $u$ is an output node; and (\textit{iii}) $v$ is an
input node.
\end{definition}
Some relations are made invisible by the access patterns and can be discarded.
\begin{definition}[Visibility]\label{def:visibility}
An input node $v_n\in\N$ in a d-graph $\tup{\N,\E}$ is \emph{visible} if there
is a sequence of arcs $u_1\arc v_1,\ldots,u_n\arc v_n$ in $\E$ such that
\begin{inparaenum}[\itshape (i)]
	\item $u_1$'s relation has no input attributes, and
	\item $v_i$'s and $u_{i+1}$'s relation are the same, for $1\leq i\leq n-1$.
\end{inparaenum}
A relation is visible if all of its input nodes are.
\end{definition}
\begin{exampleCont}{\ref{ex:answerability}}
Consider the KQ and the schemas from Example~\ref{ex:answerability}.
The d-graphs $\dgraph{q}{\SP_1}$ and $\dgraph{q}{\SP_2}$ are shown in Figures~\ref{fig:answerability1} and~Figure~\ref{fig:answerability2}, respectively.
\end{exampleCont}
Algorithm~\ref{alg:answerability} efficiently checks answerability of a KQ $q$ by means of compatibility with a schema in which all 
non-visible
 relations have been eliminated.
\begin{algorithm}[th]
\scalebox{.82}{
		\begin{minipage}{1.33\textwidth}
				\begin{compactenum}
				    \item[Input:] \emph{Schema $\SP$, KQ $q=\{k_1,\ldots,k_{|q|}\}$}
				    \item[Output:] \emph{$\codetrue$ if $q$ is answerable against $\SP$, $\codefalse$ otherwise}
					\item $\S':=$ largest subset of $\S$ s.t. all relations in $\S'^{\patternSet}_q$ are visible
					\item $\codereturn$ $\compatible(q,\S')$
				\end{compactenum}
        \end{minipage}
        }
	\caption{Checking answerability ($\answerable(q,\SP)$)}
	\label{alg:answerability}
\end{algorithm}
\begin{proposition}\label{pro:answerability-complexity}
Algorithm~\ref{alg:answerability} correctly checks answerability in PTIME.
\end{proposition}
\begin{example}\label{ex:nonanswerable}
Consider a KQ $q=\{a\colon A,c\colon C\}$ and a schema $\SP = \{r(\inputmode{A},B), s(\inputmode{C},D), u(B,D,\inputmode{E})\}$.
Relation $u$ is not visible in $\dgraph{q}{\SP}$ (Figure~\ref{fig:nonanswerable}).
According to Algorithm~\ref{alg:answerability}, $q$ is not answerable in $\SP$, since $q$ is not compatible with schema $\{r(A,B), s(C,D)\}$ (i.e., $\S$ without relation $u$).
\end{example}

\section{Extracting query answers}
\label{sec:extraction}

\vspace{-0.2cm}
Once a KQ is known to be answerable, we need to devise a strategy to extract the corresponding answers.
The simplest approach, pursued in~\cite{CMT15}, consists in extracting the entire reachable portion, by making all possible accesses with all values (either known from the KQ or, in turn, extracted via some access), and then ``carving'' out answers from it.
Such an approach may perform a lot of unnecessary accesses.
Note also that, consistently with Problem~\ref{prob:problem}, users often simply need to quickly obtain just \emph{one} answer (if any), and prefer to save potentially costly accesses. In the following, we therefore tackle Problem~\ref{prob:problem} while aiming to avoid unnecessary accesses. We still assume, for now, that the domains of the keywords are known in advance.

\paragraph{\bf Minimality of query plans.}

Ideally, given an answerable KQ $q$ and a schema $\SP$, we would like to be able to devise a strategy that always performs a minimal amount of accesses to return an answer to $q$ against every possible instance $\I$ over $\SP$.
Let $\ProgramSet$ be the set of \emph{query plans} for $q$ and $\SP$, i.e., all the deterministic programs that, for any instance $\I$ of $\SP$, compute an answer to $q$ against $\I$ by performing accesses to the relations in $\SP$. Let $\Accesses{\Program}{\I}$ indicate the set of accesses made by query plan $\Program\in\ProgramSet$ to obtain an answer to $q$. A query plan 
$\Program\in\ProgramSet$ is
 \emph{minimal} if, for every instance $\I$ over $\SP$ and for every other query plan $\Program'\in\ProgramSet$, we have $|\Accesses{\Program}{\I}|\leq|\Accesses{\Program'}{\I}|$.
Unfortunately, minimality is, in general, unattainable.
\begin{proposition}\label{pro:no-minimality}
	There are KQs that admit no minimal query plan.
\end{proposition}
With this negative result, our best hope is to attain a weaker form of minimality. 
A query plan $\Program$ for $q$ and $\SP$ \emph{dominates} $\Program'$ if 
for every instance $\I$ over $\SP$ we have
$|\Accesses{\Program}{\I}|\leq|\Accesses{\Program'}{\I}|$
and there is an instance $\I'$ over $\SP$ such that
$|\Accesses{\Program}{\I'}|<|\Accesses{\Program'}{\I'}|$;
$\Program$ is \emph{weakly minimal} if no query plan $\Program'\in\ProgramSet$ dominates it.

In the following, we will illustrate how to build a weakly minimal query plan over examples of increasing complexity.
We do this by resorting to the notion of access defined in Section~\ref{sec:problem} and to standard relational algebra operations.

\paragraph{\bf Queries with a single keyword.}
The simplest case regards KQs with a single keyword. For such queries, each answer, if any exists, consists of exactly one tuple and is therefore optimal.
We should therefore look for an access path ending with such a relation.
\begin{example}\label{ex:simplest-extraction}
Consider the KQ $q=\{a\colon A\}$ and the schema $\SP = \{r(\inputmode{A},B), s(\inputmode{B},C)\}$.
There is only one visible relation ($r$) in $\SP_q$ containing the domain of keyword $a$. It therefore suffices to extract tuples from $r$. Since we only know a value ($a$) to use as a binding for $r$ and there is no way to extract more values from abstract domain $A$, the best thing we can do, on any instance $\I$, is to access $r$ with binding $\tup{a}$, i.e., to use the access path $\access{\tup{a}}{r}{\T_r}$. Each of the resulting tuples in $\T_r$, if any, will be an optimal answer to $q$. The overall cost is $1$ access to $r$.
There is no need, of course, to access $s$ with the values in $\proj_B(s)$.
\end{example}

In some cases there may be more than one kind of access path to obtain an answer. We extract values starting with the shortest paths so as to try to reduce the number of accesses.
\begin{example}\label{ex:simple-extraction-with-choice}
Consider now schema ${\S'^{\patternSet}} = \{r(\inputmode{A},B), s(C), u(\inputmode{C},A), v(\inputmode{D},A)\}$.
There are now two visible relations ($r$ and $u$) in ${\S'^{\patternSet}_q}$ containing the domain of keyword $a$. We can either directly access $r$ with $\tup{a}$ or we can access $s$, thus retrieving values for domain $C$, and then use these values to access $u$.
Since, on any instance $\I$, any access path in
$\accessPathSet=\{\access{\tup{}}{s}{\T_s}\access{\binding}{u}{\T_u} \,\,|\,\, \binding\in\T_s\}$
requires two accesses to find an answer, we start with $\access{\tup{a}}{r}{\T_r}$. Only if $\T_r=\emptyset$ do we use a path in $\accessPathSet$ to check whether an answer exists. Again, all answers found in this way are necessarily optimal. The overall cost in the worst case is $1$ access to $r$, $1$ to $s$ and $|s|$ 
to $u$.
\end{example}

\paragraph{\bf Queries with more than one keyword.}
When a KQ involves more than one keyword, we must find appropriate access paths connecting them.
\begin{example}\label{ex:forced-chain-extraction}
Consider the schema $\SP = \{r(\inputmode{A},B), s(\inputmode{B},C)\}$ and the KQ $q=\{a,c\}$.
Suitable access paths are of the form$\access{\tup{a}}{r}{\T_r}\access{\binding}{s}{\T_s}$ with $\binding\in\proj_B(\T_r)$. Every set of tuples in $\{\{t_r,t_s\}|t_r \join t_s \in \sel_{C=c}(\T_r\join\T_s)\}$ is an optimal answer to $q$.
The overall cost in the worst case is $1$ access to $r$ and $|\proj_B(\sel_{A=a}(r))|$ to $s$.
\end{example}

The case of Example~\ref{ex:forced-chain-extraction} features a single kind of access path as well as the guarantee that a non-empty extraction of the keywords necessarily contains an answer. This does not hold when there is a relation with at least two 
output arguments.
In such cases, we use the $\peel$ procedure (Algorithm~\ref{alg:peel}) to make sure that the keywords are connected and to remove redundant tuples.
\vspace{-0.5cm}
\begin{algorithm}[th]
\scalebox{.82}{
		\begin{minipage}{1.33\textwidth}
				\begin{compactenum}
				    \item[Input:] \emph{Set of tuples $\T$, KQ $q$}
				    \item[Output:] \emph{Answer to $q$ against $\T$ or $\codenil$ if no answer}
					\item $\codeif$ $\lnot\connected(q,\T)$ $\codethen$ $\codereturn$ $\codenil$
				    \item $\codelet$ $\T':=\T$ and $\T'':=\emptyset$
					\item $\codewhile$ $\T''\neq\T'$
				    \item \ind $\codelet$ $\T'' := \T'$
				    \item\label{algline:foreach} \ind $\codeforeach$ $t\in \T''$ $\codeif$ $\connected(q,\T'\setminus\{t\})$ $\codethen$ $\codelet$ $\T' := \T'\setminus\{t\}$
					\item $\codereturn$ $\T'$
				\end{compactenum}
        \end{minipage}
        }
	\caption{Removing tuples to obtain an answer ($\peel(\T, q)$)}
	\label{alg:peel}
\end{algorithm}
\vspace{-0.5cm}
\begin{example}\label{ex:self-loop-extraction}
Consider the schema $\SP = \{r(A,B), s(\inputmode{B},C)\}$ and the KQ $q = \{a,c\}$.
A suitable query plan, essentially corresponding to the access paths $\access{\tup{}}{r}{\T_r}\access{\binding}{s}{\T_s}$, with $\binding\in\proj_B(\T_r)$, but avoiding useless accesses and computations, is shown in Figure~\ref{fig:query-plan-self-loop-extraction}. 
Consider the instance $\I = \{r(a, b), r(a_1, b), r(a_1, b_1), r(a_2,b_2), s(b_1, c), s(b_3,c_1)\}$ over $\SP$. With the query plan of Figure~\ref{fig:query-plan-self-loop-extraction} we extract $\T_r \cup \T_s= \I\setminus\{s(b_3,c_1)\}$, with $1$ access to $r$ and $3$ to $s$, and the returned (optimal) answer is $\I\setminus\{s(b_3,c_1),r(a_2,b_2)\}$.
\end{example}
\vspace{-0.5cm}
\begin{figure}[th]
\scalebox{.82}{
		\begin{minipage}{1.33\textwidth}
\begin{compactenum}
\item $\access{\tup{}}{r}{\T_r}$
\item $\codeif$ $\tup{a} \not\in \proj_A(\T_r)$ $\codethen$ $\codereturn$ $\codenil$
\item $\codeforeach$ $\binding\in\proj_B(\T_r)$
\item \ind $\access{\binding}{s}{\T_s}$
\item \ind $\codeif$ $\tup{c} \in \proj_C(\T_s)$ $\codethen$ $\codereturn$ $\peel(\T_r \cup \T_s,\{a,c\})$
\item $\codereturn$ $\codenil$
\end{compactenum}
        \end{minipage}
        }
	\caption{Query plan for Example~\ref{ex:self-loop-extraction}}
	\label{fig:query-plan-self-loop-extraction}
\end{figure}
\vspace{-0.2cm}
Determining an optimal answer from the reachable portion corresponds to finding a Steiner tree of its join graph~\cite{XuQC10}, i.e., a minimal-weight subtree of this graph involving a subset of its nodes. An efficient method for solving this problem in the context of keyword search over structured data is presented in~\cite{KiSa06}, where a \emph{$q$-fragment} can model our notion of answer. Yet, when optimality is not required, a simple technique (quadratic in the size of $\I$) to obtain an answer, shown in Algorithm~\ref{alg:peel}, consists in trying to remove any tuple from the set as long as the set after the removal contains all the keywords and remains connected.
 The $\connected$ test can be run in linear time 
  (the problem is in \textsc{logspace}~\cite{DBLP:journals/jacm/Reingold08}).
The set returned by Algorithm~\ref{alg:peel} depends on the
 order in which the tuples are enumerated (line~\ref{algline:foreach}), but this is irrelevant if we do not need an optimal answer.

The most complex scenarios involve schemas whose d-graph allows a cyclic chain of relations feeding values to other relations, which in turn determine extractions with access paths whose length cannot be bounded a priori.
\begin{example}\label{ex:cycle-extraction}
Consider schema $\SP = \{r(\inputmode{A},B), s(\inputmode{B},C,A)\}$ and KQ $q = \{a,c\}$. The d-graph $\dgraph{q}{\SP}$ (Figure~\ref{fig:cyclic-dgraph}) clearly shows a cyclic chain of relations providing values to one another (from $r$ to $s$, then back to $r$).
The best strategy is as follows.
Any access path of the form $\access{\tup{a}}{r}{\T_r}\access{\binding}{s}{\T_s}$, with $\binding\in\proj_B(\T_r)$, is the shortest candidate to extract an answer: the first extraction (using the first value in $\proj_B(\T_r)$) requires two accesses (one to $r$ and one to $s$), while all subsequent extractions (with the other values in $\proj_B(\T_r)$) require one more access to $s$. Any answer found this way is necessarily optimal.
However, longer access paths may be used to find answers.
In particular, any path attempted so far may be extended by traversing once more $r$ and then $s$ with the values obtained previously. Let $\T^{*}_s$ be the set of all the tuples extracted from $s$ so far; this new round will then consist in trying any of the access paths of the form $\access{\binding_r}{r}{\T_r}\access{\binding_s}{s}{\T_s}$, with $\binding_r\in\proj_A(\T^{*}_s)$ and $\binding_s\in\proj_B(\T_r)$. So, we first access $r$ with the first value for $\binding_r$, then $s$ with the first value for $\binding_s$
(with
 two more accesses). If no answer is found, we access $s$ with the next value for $\binding_s$ 
 (with
  one more access), and proceed this way until either an answer is found or there are no more values to try for $\binding_s$. At this point, we access $r$ with the next value for $\binding_r$ and $s$ with the first corresponding value for $\binding_s$ 
(with
two more accesses), and continue in a similar 
fashion.
When all bindings are tried, we extend again the previous access paths by traversing once more $r$ and then $s$. We proceed until either an answer is found or no new values are discovered (and thus no new accesses are possible).
\end{example}

Note that even if we explore access paths sortedly (by increasing length), we have in general no guarantee to find the optimal answers first.
\begin{exampleCont}{\ref{ex:cycle-extraction}}
Consider the KQ and the schema from Example~\ref{ex:cycle-extraction} 
and the instance $\I=\{r(a,b_1),s(b_1,c_1,a_1),r(a_1,b_2),s(b_2,c,a_2),r(a_2,b_3),s(b_3,c,a)\}$.
The access path $\access{\tup{a}}{r}{\{r(a,b_1)\}}\access{\tup{b_1}}{s}{\{s(b_1,c_1,a_1)\}}\access{\tup{a_1}}{r}{\{r(a_1,b_2)\}}\access{\tup{b_2}}{s}{\{s(b_2,c,a_2)\}}$ clearly determines an answer consisting of the $4$ tuples $\{r(a,b_1),s(b_1,c_1,a_1),r(a_1,b_2),s(b_2,c,a_2)\}$, obtained with $4$ accesses (no alternative access paths were possible so far). However, adding the accesses in $\access{\tup{a_2}}{r}{\{r(a_2,b_3)\}}\access{\tup{b_3}}{s}{\{s(b_3,c,a)\}}$ determines the answer $\{s(b_3,c,a)\}$, of size 1, which is optimal.
\end{exampleCont}

\section{Query plan generation}
\label{sec:plan-generation}

In this section we discuss the generation of a query plan similar to those shown in the previous section.
To this end, we first identify those relations in the schema that are potentially useful for retrieving an answer, i.e., they may provide either a keyword or values that, in turn, allow extracting a keyword.

\begin{definition}\label{def:useful}
Let $q$ be a KQ, $\SP$ a schema, $\SJ$ the schema join graph of $\S$, and $N$ a node in $\SJ$ whose relation $r$ is visible in $\GqS$. Node $N$ is said to be \emph{useful} for $q$ and $\SP$ if $r$ has an attribute $A$ such that either 
\begin{inparaenum}[\itshape (i)]
	\item $\exists k\in q:\dom{k}=\dom{A}$, or
	\item $A$ is output and there is a useful node $N'$ with an input attribute $A'$ such that $\dom{A}=\dom{A'}$.
\end{inparaenum}
\end{definition}

We now introduce the notion of witness, as a means to represent access paths.

\begin{definition}[Witness]\label{def:witness}
Let $q=\{k_1,\ldots,k_{|q|}\}$ be a KQ, $\SP$ a schema, and
$N_1,\ldots,N_{|q|}$ be nodes in the schema join graph $\SJ$ of $\S$ such that $N_i$'s relation has an attribute with domain $\dom{k_i}$, for $1\leq i\leq |q|$.
Let $\P$ be a path on $\SJ$ traversing all the nodes $N_1,\ldots,N_{|q|}$ and ending with $N_{|q|}$.
Let $\S'$ be the subset of $\S$ containing only the relations occurring in $\P$.
Path $\P$ is a \emph{witness} for $q$ in $\SP$ if all its nodes are useful for $q$ and $\S'$.
\end{definition}

The idea is to map each witness to an access path and each node in a witness to a relation in the schema. However, there may be infinitely many witnesses due to cycles in the schema join graph.
For this reason, Algorithm~\ref{alg:extraction} takes care of enumerating all witnesses sorted by the minimum number of accesses required to access all relations in the witness, thus finding a possible answer.
We first check whether the KQ is answerable (line~\ref{algline:answerability-check}), then we initialize 
some containers for access results (line~\ref{algline:init-empty-containers}) and domain values including the keywords (line~\ref{algline:add-keywords}). In the main extraction cycle, we select the most promising witness, i.e., the one that requires the least number of accesses to complete (line~\ref{algline:next-witness}), and then we traverse it in all possible ways in a depth-first fashion with the values we know (line~\ref{algline:traverse-path}).
For pseudocode simplicity, if an answer is found, it will be returned directly within the $\followPath$ subroutine (similarly if no answer are possible) with the $\codeexit$ instruction (all exit points are boxed for visual clarity); else we keep going as long as there are useful extractions to be done (line~\ref{algline:while-useful}).
The traversal of the witness takes place in the $\followPath$ subroutine as shown in Example~\ref{ex:cycle-extraction}.
For each binding formable with the known values (line~\ref{algline:for-every-binding}), we access the first relation in the witness, and take care of not making the access if it was already made (line~\ref{algline:access}).
With that, we update our knowledge of tuples (line~\ref{algline:update-tuples}) and values (line~\ref{algline:update-values}) and check if an answer was extracted, and in that case stop the traversal, exiting from $\extract$ with the found answer (line~\ref{algline:check-answer}).
We stop the traversal also as soon as we know that no answer can ever be found (line~\ref{algline:no-answer-possible}). Otherwise we continue the traversal (line~\ref{algline:continue-path}).
When all bindings are tried, the subroutine returns the control to its caller, thus backtracking to the previous choice.
Note that no answer can be extracted when relation $r$ contains an attribute with the domain of a keyword $k$, but $k$ does not occur in $\T_r$ and $\dom{k}$ does not occur in any later node in $\P$ or in any other unexplored witness.

\newcounter{mycounter}
\begin{algorithm}[t]
\scalebox{.82}{
		\hspace{-.75cm}\begin{minipage}{1.33\textwidth}
				\begin{compactenum}
				    \item[Input:] \emph{Schema $\SP$, KQ $q$, instance $\I$}
				    \item[Output:] \emph{An answer to $q$ against $\I$ ($\codenil$ if no answer exists)}
				    \item[Bookkeeping:] \emph{Tuple containers $\T_r$ $\forall r\in\S$, value containers $\T_D$ $\forall$ domain $D$ in $\S$}
					\item\label{algline:answerability-check} $\codeif$ $\lnot \answerable(q,\SP)$ $\codethen$ $\codeexit$ with \fbox{result $\codenil$}
					\item\label{algline:init-empty-containers} $\T_r:=\emptyset$ $\forall r\in\S$
					\item\label{algline:add-keywords} $\T_D:=\{k\mid k\in q \land \dom{k}=D\}$ $\forall$ domain $D$ occurring in $\S$
					\item\label{algline:while-useful} $\codewhile$ there is a witness for $q$ in $\SP$ with something to extract
					\item\label{algline:next-witness} \ind $\P:=$ the next witness for $q$ in $\SP$ by number of accesses to be made
					\item\label{algline:traverse-path} \ind $\followPath$($\P$) \quad \emph{// NB: this may cause exiting the main procedure with a result}
					\item $\codeexit$ with \fbox{result $\codenil$} \quad \emph{// no answer was found}
\setcounter{mycounter}{\theenumi}
				\end{compactenum}
				\hdashrule[9pt][x]{14.6cm}{0.5pt}{1pt 1pt}\vspace{-13pt}
				\begin{compactenum}\setcounter{enumi}{\themycounter}
				    \item[Subroutine:] $\followPath(\P)$, \emph{$\P$ is a path }
					\item $\codeif$ $\P$ is empty $\codethen$ $\codereturn$ \quad \emph{// path entirely explored}
					\item $r:=$ relation of the first node in $\P$
					\item\label{algline:for-every-binding} $\codeforeach$ binding $\binding$ for $r$ formable with values in $\bigcup_{D}\T_D$
					\item\label{algline:access} \ind $\codeif$ $\binding$ is a new binding for $r$ $\codethen$ $\access{\binding}{r}{\T^{+}_r}$ $\codeelse$ $\T_r^{+}=\emptyset$
					\item\label{algline:update-tuples} \ind $\T_r := \T_r \cup \T_r^{+}$ \quad \emph{// updating knowledge of $r$}
					\item\label{algline:update-values} \ind $\codeforeach$ domain $D$ in $r$ \quad $\T_D := \T_D \cup \proj_{D}(\T_r^{+})$ \quad \emph{// updating knowledge of $D$}
					\item\label{algline:check-answer} \ind $\codeif$ $\A:=\peel(\bigcup_{r\in\S}\T_r,q)\neq\codenil$ $\codethen$ $\codeexit$ everything with \fbox{result $\A$} \, \emph{// if an answer exists, stop}
					\item\label{algline:no-answer-possible} \ind $\codeif$ no answer possible $\codethen$   $\codeexit$ everything with \fbox{result $\codenil$}  \quad \emph{// if no answer may exist, stop}
					\item\label{algline:continue-path} \ind $\followPath($path $\P$ without the first node$)$ \quad \emph{// continuing extraction}
				\end{compactenum}
        \end{minipage}
        }
	\caption{Query answer extraction ($\extract(q,\SP,\I)$)}
	\label{alg:extraction}
\end{algorithm}

Note that the stopping condition in line~\ref{algline:while-useful} is met when all the bindings formable with the known values have been attempted to access all the relations of the useful nodes.
Enumerating the witnesses in increasing order of accesses to be made (line~\ref{algline:next-witness}) is key in order to achieve weak minimality. This is similar to enumerating by length of the witness, with the proviso that one may traverse the witness with fewer accesses than nodes, since the same relation may occur more than once in the witness or may have been previously accessed.

\begin{proposition}\label{pro:weak-minimality}
	Algorithm~\ref{alg:extraction} correctly computes the answer to a KQ and $\extract(q,\SP,\cdot)$ is a weakly minimal query plan for $q$ and $\SP$.
\end{proposition}

Let us finally assume that the domains of the keywords are not known in advance.
We can adopt a strategy similar to Algorithm~\ref{alg:extraction}, with the difference that, initially, unlike what is done in line~\ref{algline:add-keywords}, the keywords are added to all the domain containers $\T_D$. Then, as soon as a tuple containing a keyword $k$ is extracted by some access, the domain $\dom{k}$ is discovered and $k$ can be removed from the containers for the other domains.

\section{Related Work}\label{sec:related}

To our knowledge, this is the first paper that proposes a comprehensive approach to the problem of querying the Deep Web using keywords. In an earlier work~\cite{CMT15}, we have just defined the problem 
and provided some preliminary insights on query processing.

The problem of query processing in the Deep Web has been widely investigated in the last years, with different approaches and under different perspectives including: data crawling~\cite{RaGa01}, integration of data sources~\cite{HeZC05}, query plan optimization~\cite{CaMa:ICDE2008}, indexing through pre-computation of forms~\cite{MAAH09}, question answering~\cite{Lehm12}, and generic structured query models~\cite{JaJa15}. However, none of them has tackled the problem that we have addressed in this paper.

The idea of querying structured data using keywords emerged more than a decade ago~\cite{AgCD02,BHNC02} as a way to provide high-level access to data and free the user from the knowledge of query languages and data organization. Since then, a lot of work has been done in this field (see, e.g.,~\cite{XuQC10} for a survey) but never in the context of the Deep Web.  
This problem has been investigated in the context of various data models: relational~\cite{KiSa06}, semi-structured~\cite{DBLP:conf/sigmod/LiOFWZ08}, XML~\cite{DBLP:conf/sigmod/GuoSBS03}, and RDF~\cite{DBLP:conf/icde/TranWRC09}. 
Within
 the relational model, the common assumption is that an answer to a keyword query is a graph of minimal size in which the nodes represent tuples, the edges represent foreign key references between them, and the keywords occur in some node of the graph~\cite{DBLP:conf/vldb/HristidisP02}. 
Our definition of query answer follows this line but it is more general, since it is only based on the presence of common values between tuples, while not forcing the presence of foreign keys.

The various approaches to keyword query answering over relational databases are commonly classified into two categories: \emph{schema-based} and \emph{schema-free}. Schema-based approaches~\cite{AgCD02,DBLP:conf/vldb/HristidisP02} make use, in a preliminary phase, of the database schema to build SQL queries that are able to return the answer. Conversely,
schema-free approaches~\cite{BHNC02,DBLP:conf/sigmod/GolenbergKS08,KiSa06,DBLP:conf/sigmod/LiOFWZ08} rely on exploration techniques over a graph-based representation of the whole database. Since the search for an optimal answer consists in finding a minimal \emph{Steiner tree} on the graph, which is known to be an NP-Complete problem~\cite{GareyGrahamJohnson}, the various proposals rely on heuristics aimed at generating approximations of Steiner trees. 
Our approach makes use of the schema of the data sources but cannot be classified in any of the approaches above since, given the access limitations, it rather relies on building a minimal query plan of accesses to the data sources.

This paper builds on prior work by integrating schema-awareness with graph-based exploration techniques. Unlike conventional methods, it explicitly models access constraints and defines a query processing framework that minimizes access costs. Similar ideas have been explored in the context of structured data and databases with known schemas~\cite{CM:ICDE2008,CaMa10,DBLP:conf/er/CaliM08,CCM:JUCS2009}. However, this work extends these principles to the Deep Web, proposing efficient query plan generation and optimization strategies for dynamically retrieved data. By addressing this gap, the paper lays the foundation for scalable and practical keyword search over inaccessible data repositories.

We observe that the access patterns used here are a form of integrity constraints that can be exploited for query optimization, schema specification and data maintenance~\cite{M:PHD2005,DBLP:conf/lopstr/ChristiansenM03,DBLP:conf/foiks/ChristiansenM04,DBLP:conf/fqas/Martinenghi04,DBLP:conf/dexa/MartinenghiC05,DBLP:journals/aai/ChristiansenM00,DBLP:conf/dexaw/DeckerM07,DBLP:conf/lpar/DeckerM06,DBLP:conf/dexaw/DeckerM06,DBLP:journals/tplp/CaliM10}, and it should be interesting to study the interaction between access patterns and integrity constraints proper.

It should also be interesting to understand whether preferences and context information --- a ubiquitous concern in databases~\cite{DBLP:journals/jacm/CiacciaMT20,DBLP:conf/fqas/MartinenghiT09} --- can be used to drive the selection of query plans so as to retrieve answers in a best-first manner, unlike the semantics-agnostic criterion used here to produce suitable answers.

Keyword searches are common components of larger pipelines involving data scraping and searching based on Machine Learning steps~\cite{DBLP:conf/fqas/Masciari09,DBLP:journals/isci/MasciariMZ14,DBLP:conf/ismis/MasciariMPS20,DBLP:conf/ideas/MasciariMPS20}, with data retrieved from sources of different nature, such as  RFID~\cite{DBLP:conf/ideas/FazzingaFMF09,DBLP:journals/tods/FazzingaFFM13}, pattern mining~\cite{DBLP:conf/ideas/MasciariGZ13}, crowdsourcing applications~\cite{DBLP:conf/socialcom/GalliFMTN12,DBLP:conf/www/BozzonCCFMT12,DBLP:conf/mmsys/LoniMGGMAMMVL13}, and streaming data~\cite{DBLP:journals/jiis/CostaMM14}, often mixing different levels of data distribution and access policies~\cite{DBLP:journals/tkde/MartinenghiT12,DBLP:conf/cikm/CiacciaM18}.

\section{Conclusions and future work}
\label{sec:conclusions}

In this paper, we have defined the problem of keyword search in the Deep Web and provided a method for query answering in this context that aims at minimizing the number of accesses to the data sources. 

We believe that several interesting issues can be
studied
 in the framework defined in this paper.
We plan, e.g., to leverage known values (besides the keywords) and ontologies to speed up the search for an optimal answer as well as to consider the case in which nodes and arcs of the join graph are weighted to model source availability and proximity, respectively.

\vspace{-0.2cm}
\bibliographystyle{plain}

\section*{Appendix}
\subsection*{Proof of Proposition~\ref{pro:compatibility-correct}}
\begin{proof}
First, note that if the domain of a keyword is not present in $\S$, the keyword may never be found, and thus the algorithm returns $\codefalse$ (line~\ref{algline:domainnotfound}). Also, note that relations of arity $1$ are irrelevant for connecting two keywords in an answer: indeed, if a keyword $k$ occurs in an atom of a unary relation and is connected to another keyword $k'$, then it must also occur in a non-unary atom. Therefore, such relations are not taken into account for checking connectedness (line~\ref{algline:nonunary}).
In the remainder of Algorithm~\ref{alg:compatibility}, $\codefalse$ is returned when the domains of two keywords in $q$ can never be connected in the schema join graph $\SJ$ (line~\ref{algline:ifnopath}), which entails that $q$ is not compatible with $\S$.
Conversely, when Algorithm~\ref{alg:compatibility} returns $\codetrue$, then, for $1\leq i\leq |q|-1$, $\SJ$ has a path (of, say, $n_i$ nodes) connecting the domain of a keyword $k_i$ in $q$ to the domain of keyword $k_{i+1}$ in $q$. Therefore, one can build a sequence of $n_i$ tuples $t_{i,1},\ldots,t_{i,n_i}$ such that $t_{i,1}$ contains keyword $k_i$, $t_{i,n_i}$ contains $k_{i+1}$, and $t_{i,j}$ shares a value with $t_{i,j+1}$ for $1\leq j< n_i$.
Note that we may also have a path of $1$ node (on, say, relation $r$) such that
\begin{inparaenum}[\itshape (i)]
	\item $\dom{k_i}=\dom{k_{i+1}}$, and
	\item only one attribute in $r$ uses domain $\dom{k_i}$.
\end{inparaenum}
In this special case, we set $n_i=2$ and we build a sequence of 2 tuples $t_{i,1},t_{i,2}$, both on $r$, connected by sharing a value on one of the other attributes in $r$ (which is necessarily non-unary).

This construction step can be repeated for every two keywords $k_i$ and $k_{i+1}$, for $1\leq i\leq |q|-1$, thus obtaining an overall set of tuples $\T=t_{1,1},\ldots,t_{1,n_1},t_{2,1},\ldots,t_{2,n_2},\ldots,t_{|q|-1,1},\ldots,t_{|q|-1,n_{|q|-1}}$. Note that the join graph of $\T$ is connected, since every sequence $t_{i,1},\ldots,t_{i,n_i}$ is connected, and every two consecutive sequences are connected, since both $t_{i,n_i}$ and $t_{i+1,1}$ share the same keyword $k_{i+1}$.
Since $\T$ is connected and contains all the keywords in $q$, $\T$ (or a subset thereof) is an answer to $q$ against an instance $\I=\T$.

All the steps in Algorithm~\ref{alg:compatibility} are polynomial in the size of the schema or of the query, and line~\ref{algline:ifnopath} amounts to checking reachability, which is again polynomial in the size of the schema.
\end{proof}

\subsection*{Proof of Proposition~\ref{pro:answerability-complexity}}
\begin{proof}
No tuple can be extracted from a non-visible relation.
Certainly, if no path in the schema join graph of $\S'$ can be found that connects two keywords in $q$, then no connected set of tuples with these keywords can be extracted.

Conversely, if such paths can be found for all pairs of keywords in $q$, then a set of tuples $\T$ providing an answer to $q$ can be formed in the same way as in the proof of Proposition~\ref{pro:compatibility-correct}. The presence of access patterns may in some case force some values in $\T$ to coincide, but this does not affect connectedness of $\T$.

Visibility can be checked in PTIME since it amounts to checking reachability (at the level of a relation) on a (dependency) graph.
Compatibility was shown to be in PTIME in the proof of Proposition~\ref{pro:compatibility-correct}.
Therefore answerability is also in PTIME.
\end{proof}

\subsection*{Proof of Proposition~\ref{pro:no-minimality}}

\begin{proof}
Consider a KQ $q=\{a\colon A, b\colon B\}$ and a schema $\SP=\{r(\inputmode{A},B), s(A,\inputmode{B})\}$.
Any query plan $\Program\in\ProgramSet$ must either first access $r$ with the binding $\tup{a}$ or access $s$ with the binding $\tup{c}$.
Now, consider the instances $\I_1=\{r(a,b)\}$ and $\I_2=\{s(a,b)\}$.
If $\Program$ accesses $r$ first, then there exists another query plan $\Program'$ such that $\Accesses{\Program'}{\I_2}\subset\Accesses{\Program}{\I_2}$, since the access to $r$ is useless to find the answer $s(a,b)$.
Similarly, if $\Program$ accesses $s$ first, then such an access is useless to find the answer $r(a,b)$ from $\I_1$.
Therefore, there is no minimal query plan for $q$ and $\SP$.
\end{proof}

\subsection*{Proof of Proposition~\ref{pro:weak-minimality}}

\begin{proof}
Correctness follows from the fact that all possible ways of extracting an answer are eventually found (by enumerating all witnesses), and by correctness of the $\peel$ procedure, which we use after each extraction to check if an answer was found.

Weak minimality follows from the fact that the strategy used in Algorithm~\ref{alg:extraction} is never dominated by another strategy. Indeed, the first witness $\P$ attempted by Algorithm~\ref{alg:extraction} is the one that requires the least amount of accesses. Take any query plan $\Program$ not attempting an access path along $\P$ as its first choice: then there would be an instance $\I$ (providing an answer that can be extracted with an access path along $\P$) in which $\Accesses{\Program}{\I}>\Accesses{\extract(q,\SP,\cdot)}{\I}$, i.e., $\Program$ would incur more accesses than Algorithm~\ref{alg:extraction}.
The same argument would apply to any query plan making the same first choice as Algorithm~\ref{alg:extraction} but a different second attempt to extract an answer, since the second witness is the second-best choice by number of accesses to be made; and so on for the following choices.
Note that two query plans attempting the same paths in the same order might still differ by the order in which the different bindings are attempted (as in the enumeration in line~\ref{algline:for-every-binding} in the $\followPath$ subroutine). However, by a symmetry argument, if there is an instance $\I$ in which Algorithm~\ref{alg:extraction} incurs more accesses than another query plan differing by the order in which the bindings are attempted, then, since these are deterministic algorithms and their choices are always the same when the knowledge is the same, there must also be an instance $\I'$ in which the opposite happens.
We can therefore conclude that any query plan making different choices than Algorithm~\ref{alg:extraction} will incur more accesses in at least one instance. Therefore Algorithm~\ref{alg:extraction} is weakly minimal.
\end{proof}

\end{document}